\documentclass[]{spie}  

\usepackage{amsmath,amsfonts,amssymb}
\usepackage{graphicx}
\usepackage[colorlinks=true, allcolors=blue]{hyperref}

\usepackage[english]{babel}
\usepackage{hyperref}
\usepackage{rotating} 
\usepackage{caption, tabularx}
\usepackage[svgnames]{xcolor}

\title{Deep learning-based super-resolution fluorescence microscopy on small datasets}
\author[1]{Varun Mannam$^*$}
\author[1,2]{Yide Zhang}
\author[1]{Xiaotong Yuan}
\author[1]{Scott Howard}

\affil[1]{Department of Electrical Engineering, University of Notre Dame, Notre Dame, IN 46556, USA}
\affil[2]{Caltech Optical Imaging Laboratory, Andrew and Peggy Cherng Department of Medical Engineering,  California Institute of Technology, Pasadena, CA 91125, USA}
\authorinfo{*e-mail: vmannam@nd.edu}

\begin{document} 
\maketitle

\begin{abstract} 
Fluorescence microscopy has enabled a dramatic development in modern biology by visualizing biological organisms with micrometer scale resolution. However, due to the diffraction limit, sub-micron/nanometer features are difficult to resolve. While various super-resolution techniques are developed to achieve nanometer-scale resolution, they often either require expensive optical setup or specialized fluorophores. In recent years, deep learning has shown potentials to reduce the technical barrier and obtain super-resolution from diffraction-limited images. For accurate results, conventional deep learning techniques require thousands of images as a training dataset. Obtaining large datasets from biological samples is not often feasible due to photobleaching of fluorophores, phototoxicity, and dynamic processes occurring within the organism. Therefore, achieving deep learning-based super-resolution using small datasets is challenging. We address this limitation with a new convolutional neural network based approach that is successfully trained with small datasets and achieves super-resolution images. We captured 750 images in total from 15 different field-of-views as the training dataset to demonstrate the technique. In each FOV, a single target image is generated using the super-resolution radial fluctuation method. As expected, this small dataset failed to produce a usable model using traditional super-resolution architecture. However, using the new approach, a network can be trained to achieve super-resolution images from this small dataset. This deep learning model can be applied to other biomedical imaging modalities such as MRI and X-ray imaging, where obtaining large training datasets is challenging.
\end{abstract}


\section{Introduction}  \label{introduction}
Fluorescence microscopy has been widely used in biological and cellular studies. However, the spatial resolution of an image generated by conventional fluorescence microscopy is limited by the light diffraction to a few hundred nanometers. The limited resolution hinders further observation and investigation of objects at a molecular scale, such as mitochondria, microtubules, nanopores, and proteins (less than a few hundred nanometers) within the cells and tissues. There are many super-resolution microscopy techniques that can overcome the diffraction limit and achieve super-resolution up to ten times greater than that in conventional microscopy techniques. Experimental methods including stimulated emission depletion (STED), structured illumination microscopy (SIM), non-linear SIM \cite{huszka2019super} can generate super-resolution images, but require additional hardware setups. Recently, our group demonstrated a new super-resolution technique called generalized stepwise optical saturation (GSOS) \cite{GSOS} to improve the image resolution by reducing the point spread function (PSF) width. Localization and statistical approaches including stochastic optical reconstruction microscopy (STORM) and photo-activated localization microscopy (PALM) \cite{huszka2019super} can also enhance the image resolution, but require special fluorophores that emit light periodically and extensive computation.  Super-resolution radial fluctuation (SRRF) \cite{SRRF_Nature} is a novel super-resolution technique used to perform super-resolution imaging. SRRF can generate images with a resolution comparable to localization approaches (around 70 nm) without the need for complicated hardware setups and special imaging conditions. Even so, it requires numerous images to be collected and long computation time.
\begin{figure}[!ht]
\centering
\includegraphics[page=1,width=1\linewidth]{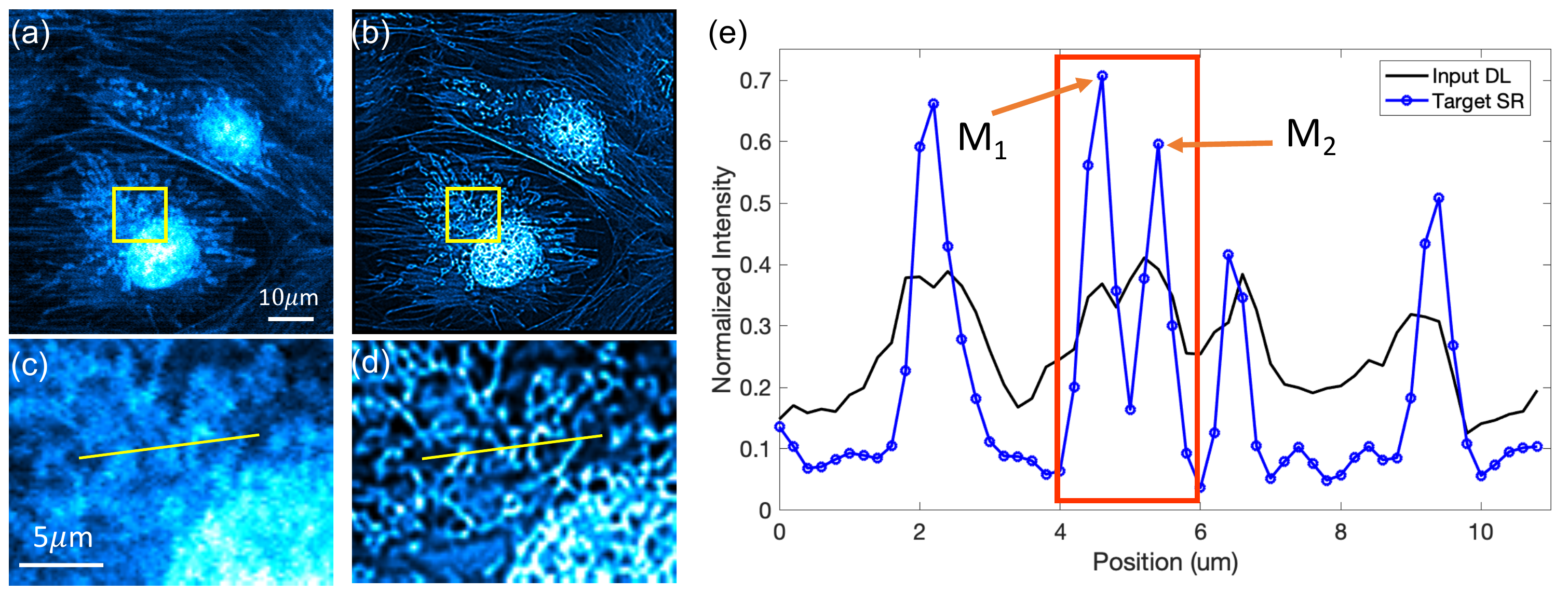}
\caption{Illustration of super-resolution in fluorescence microscopy. BPAE cells labelled with MitoTracker Red CMXRos (mitochondria), Alexa Fluor 488 phalloidin (F-actin), and DAPI (nuclei) acquired with our custom-built two-photon FLIM system \cite{instantFLIM}. Diffraction-limited image (a), super-resolution image generated using the SRRF method \cite{gustafsson2016fast} (b), a small field-of-view (FOV) from the diffraction-limited and super-resolution images in (c) and (d) respectively. (e) Line plots of normalized intensity of diffraction-limited and super-resolution images. $M_1$ and $M_2$ indicates two fluorophores. Excitation power: 3mW.} \label{sr_fig1}
\end{figure}

Fig.~\ref{sr_fig1}(a) and (b) show both the diffraction-limited (DL) image and the super-resolution (SR) image generated using the SRRF method. The images show BPAE cells labeled with MitoTracker Red CMXRos (mitochondria), Alexa Fluor 488 phalloidin (F-actin), and DAPI (nuclei) and are acquired with our custom-built instant FLIM system \cite{instantFLIM}. The selected field of view (FOV) (shown in yellow color) of diffraction-limited and super-resolution images are shown in (c) and (d), respectively. A line is selected (shown in yellow color) to show the two fluorophores ($M_1$ and $M_2$) of the diffraction-limited and super-resolution images' line plot. The fluorophores cannot be well differentiated in the diffraction-limited image, whereas the super-resolution image clearly shows these two fluorophores.  

Machine learning (ML) has been catching attention by its fast processing speed and wide applications, such as image classification, denoising, and segmentation \cite{mannam2020machine}. Recently, advanced ML models have been demonstrated to estimate super-resolution images from diffraction-limited ones. For instance, applying ML models, specifically deep neural networks (DNNs), to generate super-resolution images is presented in \cite{DeepSTORM}. The network's performance is evaluated by comparing the estimated model output images with the target images generated from a super-resolution technique (STORM). Although the feasibility and efficiency of this Deep-STORM approach have already been validated, special experimental conditions and heavy data processing are still needed for the target generation. Similarly, a few ML-based methods demonstrated to achieve super-resolution from diffraction-limited images are provided in Table.~\ref{ML_compare}. We broadly classify these ML models into two categories: first, Fully Convolutional (FC) layers containing either encoder and decoder blocks \cite{mannam2020performance} or residual connection block; second, generative adversarial networks (GANs) containing generator and discriminator blocks \cite{osokin2017gans}. All of these ML-based methods are data-driven and require a massive training dataset containing thousands of images. However, acquiring a massive dataset is challenging for \textit{in vivo} imaging where biological samples are prone to photobleaching and could show fast dynamics. Similarly, other biomedical imaging modalities such as MRI and X-ray are challenging for obtaining large training datasets. Hence, there is a strong demand for a ML model that provides super-resolution images by training with a small dataset (typically less than 1000 images). In the next two sections, the ML model architecture and the network configuration are provided. Finally, the estimated super-resolution image from the diffraction-limited image is provided with various network configurations, and it is compared with the target super-resolution image qualitatively.

\section{Methods} \label{methods}
This section introduces our proposed ML model (DenseED) approach to achieve super-resolution images when training with a small dataset. Developed for physical systems and computer vision tasks, DenseED \cite{zhu2018bayesian} is the state-of-the-art CNN architecture due to its backbone of dense layers, which passes the extracted features from the previous layer to all next layers in a feed-forward fashion. To understand this dense layer, we need to understand the convolutional layers' regular operation in fully convolutional layers. This paper also shows how to utilize these blocks to build our super-resolution ML model that works with a small dataset.

\begin{figure}[!t]
\centering
\includegraphics[page=2,width=1\linewidth]{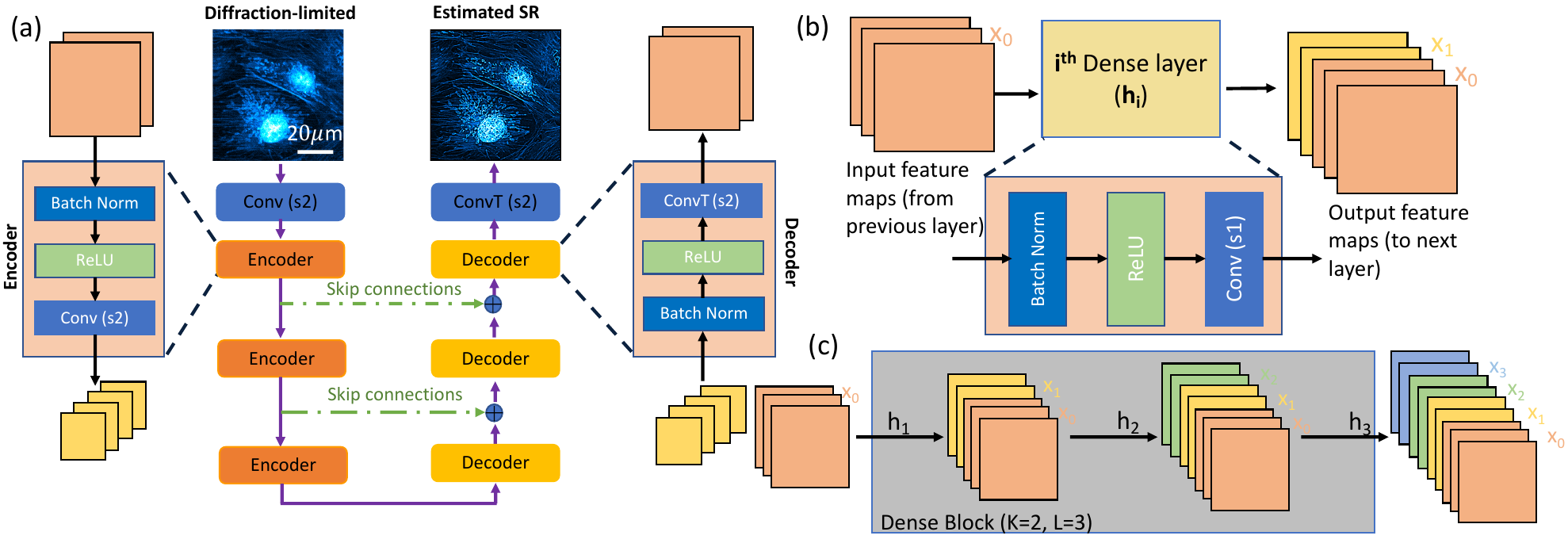}
\caption{Block diagram of Fully convolutional layers with skip connections (a), including the encoder and decoder blocks. (a) Shows the Auto encoder (AE) and U-Net architectures without and with skip connections respectively. (b) shows the Dense layer connections that provide two feature maps as the output, and (c) Dense Block consists of three dense layers with growth rate of two, i.e., each dense layer provides two feature maps as the output.} \label{sr_fig2}
\end{figure}

The fully convolutional layers \cite{long2015fully} are used for pixel-wise prediction, e.g. semantic segmentation, image denoising, super-resolution and low-dose CT X-ray reconstruction. Fig.~\ref{sr_fig2}(a) shows the fully convolutional layers with encoding and decoding blocks (without skip connections). Conv(s2) indicates the convolutional layer with a stride of 2. The convolutional layer contains the input image convolved with a kernel (filter) that extracts particular features from the input images (for example, edges, background, objects with different shapes). Here the number of filters used in the convolutional layer is called ``number of feature maps" and the output of the convolution indicates the ``feature map" with its dimension as ``feature map size". Typically an encoding block contains the convolutional layer with double feature maps and half of the feature map size. 
Fig.~\ref{sr_fig2}(a) encoder block shows the input with two feature maps of size $s$ and the output of encoding block results in four feature maps of the size of $s/2$. In this way, we are selecting only the essential features as the output of the encoder block. The decoder block works exactly opposite to the encoder block. The decoder block's output reduces the number of feature maps to half and double the feature map size. To extract the complex features, more encoder and decoder blocks are required. However, after some encoding blocks, the feature map reaches a minimum image dimension, and it cannot be restored by using decoder blocks. In other words, coarse features are not passed through the decoder blocks. This minimum image dimension of the encoder is called the ``latent space". Also, as the number of encoder and decoder blocks increases, the number of filter parameters to estimate increases exponentially, which are parameter inefficient.

Some cases require more encoder and decoder blocks. As the number of encoder and decoder blocks increases, the feature map size is reduced and the essential features are lost. Introducing the ``skip connections" between encoder and decoder blocks to pass finer features to the decoder blocks. This modified FC layers architecture is called ``U-Nets" \cite{ronneberger2015u, mannam2020instant}. This is shown in Fig.~\ref{sr_fig2}(a) with green arrows and $\oplus$ indicates the concatenation of features from encoder block and the output of previous decoder block. Another way of passing features from one layer to the next layer is using dense layers. Dense layers \cite{huang2017densely, jegou2017one} are used to create dense connections between all layers to improve the information (gradient) flow through the complete ML model for better parameter efficiency. Fig.~\ref{sr_fig2}(b) shows the dense layer connection for $i^{th}$ layer with input feature maps of $x_0$ (output of the previous layer) and passed through the dense layer with output feature maps of $x_1$; total feature maps are the concatenation of input and output feature maps [$x_0$, $x_1$]. In the dense layer, the convolution operation is performed with a stride of 1. Fig.~\ref{sr_fig2}(c) shows a dense block with three dense layers of each layer that provides two feature maps as output. In the dense block, the dense layer establishes connections from the previous layer to all subsequent layers. To put it in another way, the input features of one layer is concatenated to this layer's output features, which serves as the input features to the next layer. Let the input has $K_0$ feature maps, and each layer of the outputs has $K$ feature maps, then the $i^{th}$ layer would have input with $K_0 + (i*K)$ feature maps, i.e., the number of feature maps in dense block grows linearly with the depth and $K$ is here referred to the growth rate.

\begin{figure}[!t]
\centering
\includegraphics[page=3,width=0.8\linewidth]{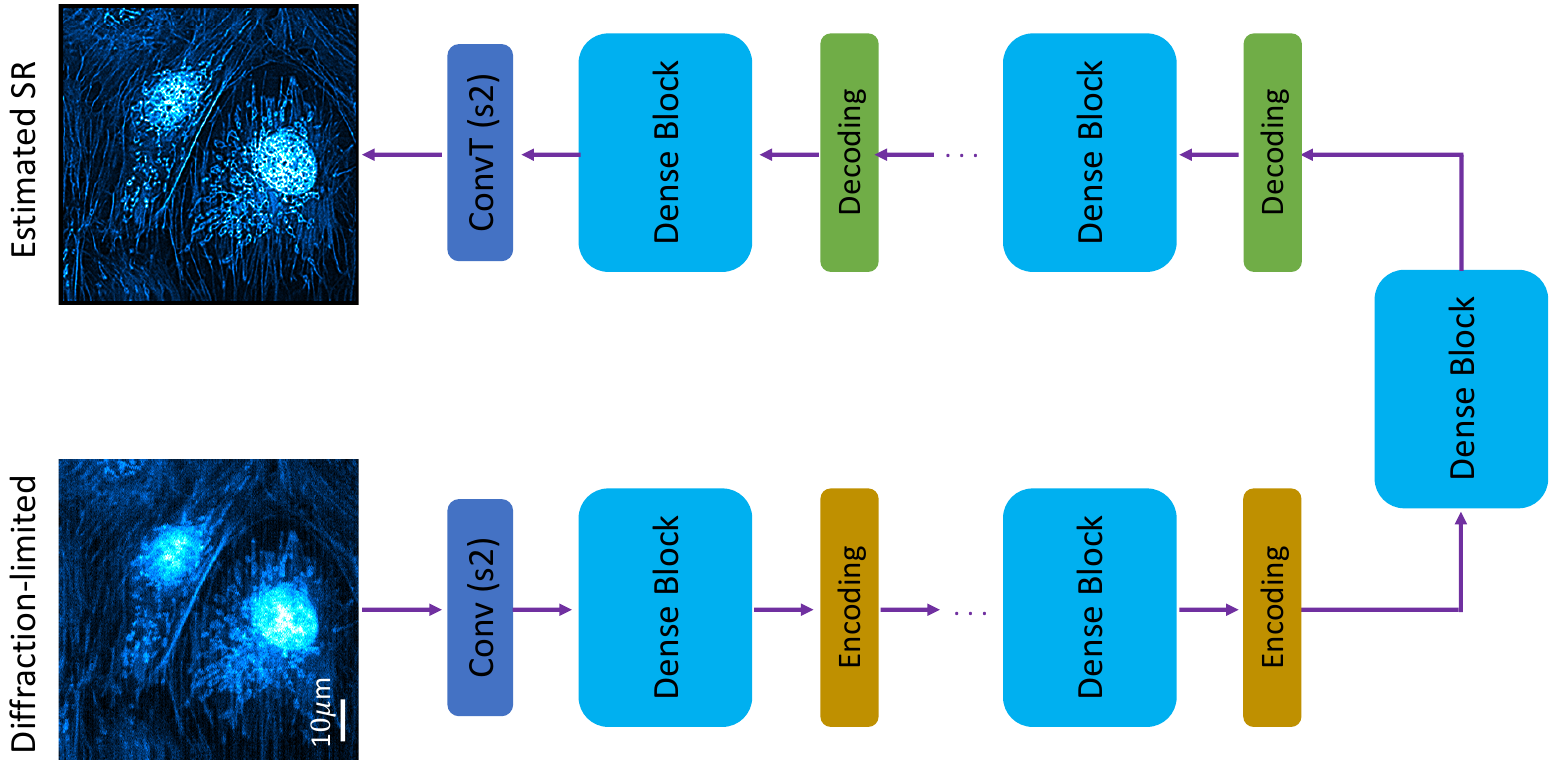}
\caption{DenseED ML model architecture shows the Dense Blocks and Encoding and decoding blocks.} \label{sr_fig3}
\end{figure}

For image regression based on the fully convolutional layer, encoding and decoding blocks are required to change the size of feature maps, making the concatenation of feature maps unfeasible across different feature map size blocks. Hence particular encoding and decoding blocks are used to solve this problem. A dense block contains multiple dense layers whose input and output feature maps are the same size. It contains two design parameters: the number of layers $L$ with the growth rate $K$ for each layer. In this work, we consider the growth rate $K$ to be a constant values for all the dense blocks. Here the encoding blocks typically half's the size of feature maps, while the decoding block doubles the feature map size. Both of the two blocks reduce the number of feature maps to half. Fig.~\ref{sr_fig3} shows the complete DenseED ML model used to generate the super-resolution images using a small dataset. Dense blocks, encoding, and decoding blocks are marked with different colors.

Different DenseED configurations and corresponding parameters are given in Appendix ~\ref{A1}. After extensive architecture search, here we presented three different DenceED configurations (DenseED - (3,6,3), DenseED - (8,8,8), DenseED - (9,18,9)) as the ML models. We choose these models with a minimal number of parameters and accurately estimate super-resolution images. Hyper-parameter search is a critical step for quick and accurate results in the deep learning area, mostly problem-specific and empirical. We run the model configurations with different hyper-parameters including learning rate, weight decay, number of output channels of the first convolution and batch size during training. Since we have a small dataset, we sliced each training image of dimension 256 $\times$ 256 into four slices so that each training image slice is 128 $\times$ 128. In this way, we increase the training dataset size with a smaller feature map size. The ML models are trained with 200 epochs so that MSE loss reaches a stable point. Table.~\ref{ML_hyper} shows the hyper-parameters used during the training of DenseED models.
\begin{table}[!ht]
\begin{tabular}{|c|c|}
\hline
\textbf{Parameter} & \textbf{Value} \\\hline
 Batch size & 4              \\\hline
Initial learning rate & 3$\mathrm{e}$-4  \\\hline
Weight decay  & 3$\mathrm{e}$-5           \\\hline
Initial number of features & 48 \\\hline
Each Dense layer output features  & 16 \\\hline           
\end{tabular}
\caption{Summary of hyper-parameters used to train the DenseED ML model configurations.}
\label{ML_hyper}
\end{table}

\section{Results}  \label{results}
\begin{figure}[!ht]
\centering
\includegraphics[page=4,width=1\linewidth]{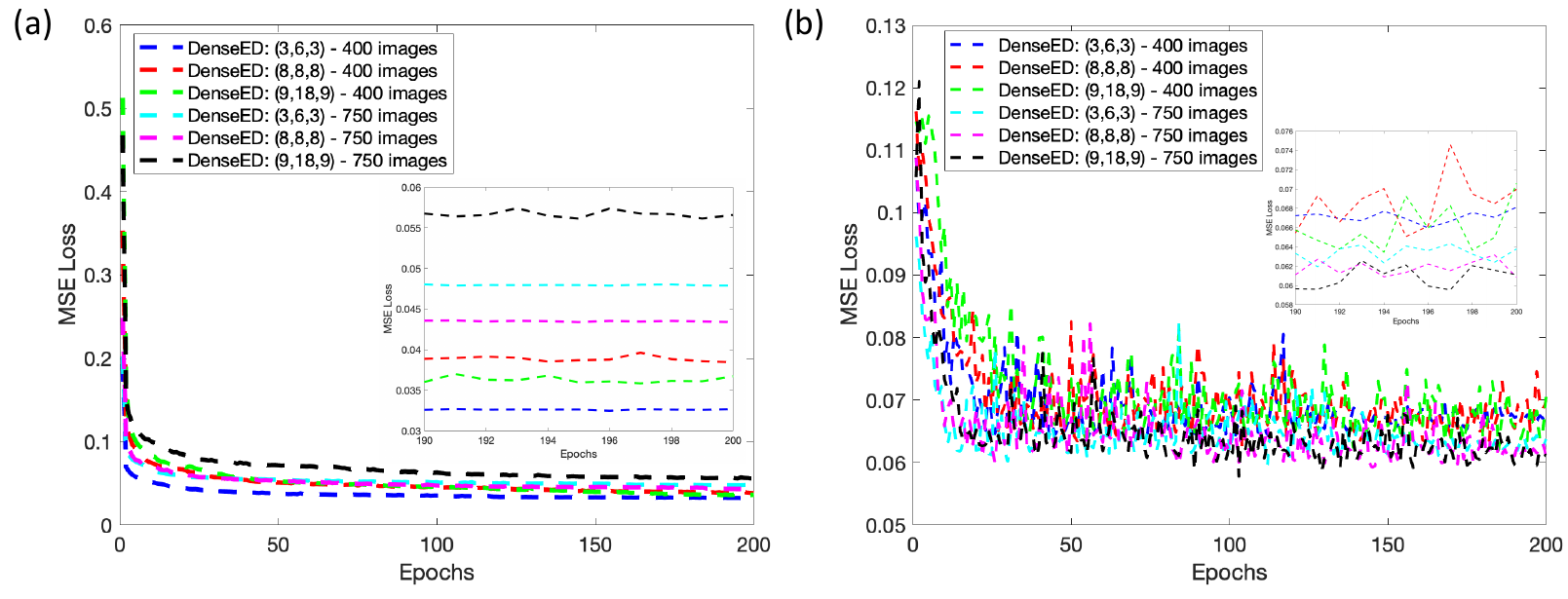}
\caption{MSE loss of (a) training and (b) testing dataset with three DenseED configurations and training dataset of 8 and 15 FOV's. Inset shows the zoomed version of the MSE loss of training and testing dataset over the last 10 epochs.} \label{sr_fig4}
\end{figure}
In this section, we show the results and the loss function, which can validate the performance of our trained ML model. First, we train the DenseED ML models with a small dataset consisting of 8 FOV's (400 images: each FOV contains 50 diffraction-limited images). During the training, we use Adam as an optimizer, and the mean square error (MSE) loss between the estimated super-resolution image and the target image as the loss function. The training image dataset contains diffraction-limited images of size 256 $\times$ 256 with a single gray channel as the input. The output super-resolution image also has a single channel. To validate the network model, we use single-FOV diffraction-limited images. We perform testing on a single image that the model never sees during the training step. Fig.~\ref{sr_fig4}(a) and (b) show the training and testing MSE loss converged after 150 epochs for the three DenseED configurations mentioned above. Later, we train the DenseED configurations with 15 FOV's training data (750 images) to see MSE loss by adding more training data, and results are included in the same figure. During the testing phase, DenseED models trained with 15 FOV's show better improvement than the models trained with only 8 FOV's.

\begin{figure}[!ht]
\centering
\includegraphics[page=5,width=1\linewidth]{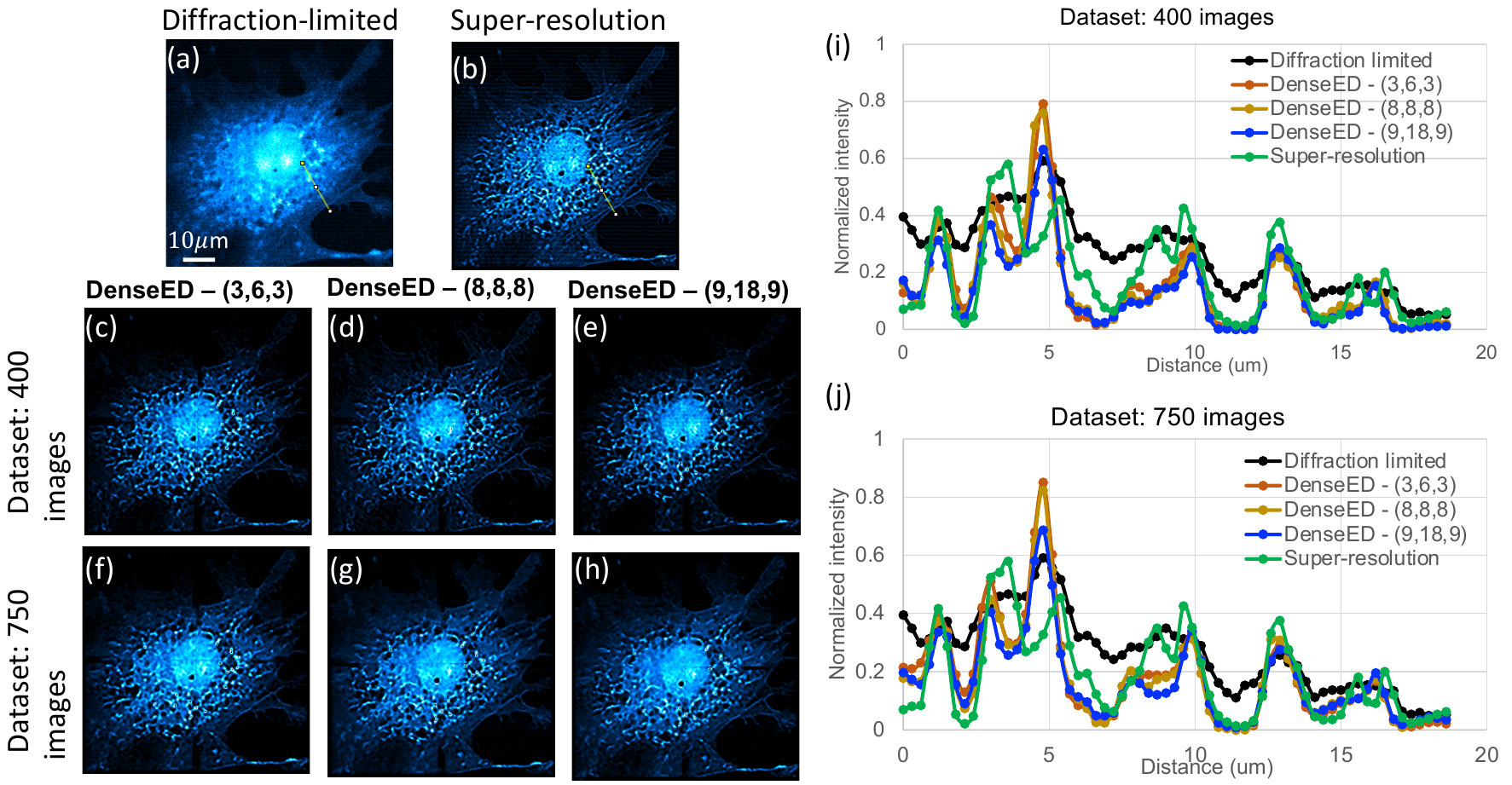}
\caption{BPAE cells diffraction-limited image (a) acquired with our custom-built two-photon microscopy and super-resolution target image generated by SRRF method (b). Estimated super-resolution images with 400 images training dataset with DenseED configurations: (3,6,3), (8,8,8) and (9,18,9) in (c), (d) and (e), respectively. Training datasets with 750 images using the above DenseED configurations are given in (f), (g) and, (h), respectively. Estimated super-resolution images line plots when trained with 400 images and 750 images are shown in (i) and (j), respectively. Power: 6 mW.} \label{sr_fig5}
\end{figure}

During the testing phase, the trained DenseED configuration's accuracy is validated by an estimated super-resolution image from a diffraction-limited image. Fig.~\ref{sr_fig5}(a) shows one of the diffraction-limited images in a testing FOV. Fig.~\ref{sr_fig5}(b) shows the target super-resolution image using the SRRF method using the testing FOV. Fig.~\ref{sr_fig5}(c), (d) and (e) shows the estimated super-resolution images from the DenseED configurations (3,6,3), (8,8,8) and (9,18,9) and trained using 8-FOV's dataset. Clearly, the estimated super-resolution image using configuration (9,18,9) shows more accurate results compared to others. Fig.~\ref{sr_fig5}(i) shows the line plot of different estimated super-resolution methods trained with 400 images. In the line plot, from 5$\mu$m to 10$\mu$m, the two different fluorophores which was hided by the diffraction-limited image are appeared in the estimated super-resolution images and the target. Similarly, Fig.~\ref{sr_fig5}(f), (g), and (i) show the estimated super-resolution images when trained in the same DenseED configurations with 750 images dataset. Also, Fig.~\ref{sr_fig5}(j) shows the line plot of different estimated super-resolution methods trained with 750 images. From these results, estimated super-resolution images are better when trained with a large training dataset. The results used in this manuscript are provided in open-source and can be accessed via GitHub \footnote{\url{https://github.com/ND-HowardGroup/SPIE-CNN-SR.git}}.

\section{Conclusion}
Machine learning has been previously demonstrated generating super-resolution from diffraction-limited images. Such approaches require thousands of training images, which is prohibitively difficult in many biological samples. We demonstrated DenseED architecture to train super-resolution convolutional neural networks using small (400/750 images with 8/15 FOV's) datasets that is impossible using conventional ML training models. We introduced our DenseED model with three different configurations and chose the best super-resolution results. The super-resolution image can also be obtained from a resolution-limited image with GANs in machine learning. We can compare DenseED results in FC layers model performance with GANs with DenseED blocks as one of the essential future directions. Also, training the ML models with more training images for the same DenseED model improves super-resolution images.

\section*{Disclosures}
\noindent The authors declare no conflicts of interest.

\section*{Funding.}
This material is based upon work supported by the National Science Foundation (NSF) under Grant No. CBET-1554516. 

\acknowledgments 
Yide Zhang’s research was supported by the Berry Family Foundation Graduate Fellowship of Advanced Diagnostics $\&$ Therapeutics (AD$\&$T), University of Notre Dame. The authors further acknowledge the Notre Dame Center for Research Computing (CRC) for providing the Nvidia GeForce GTX 1080-Ti GPU resources for training the Super-resolution (SR) Fluorescence Microscopy dataset\footnote{\url{https://curate.nd.edu/show/5h73pv66g4s}} in Pytorch.

\appendix
\section{DenseED configurations and parameters} \label{A1}
Table.~\ref{models_compare} shows the different DenseED configurations and corresponding convolution layers, parameters to tune, and maximum feature size for the configuration.
\begin{table}[!ht]
\centering
\begin{tabular}{|p{2cm}|p{3cm}|p{2.5cm}|p{2cm}|p{3cm}|}
\hline
ML Model & Configuration                        & Convolutional layers & Number of parameters & Maximum feature maps \\ \hline
DenseED  & (1,1,1)                              & 10                   & 36572                & 64                   \\ \hline
DenseED  & (2,2,2)                              & 13                   & 80702                & 72                   \\ \hline
DenseED  & (3,3,3)                              & 16                   & 143040               & 96                   \\ \hline
DenseED  & (4,4,4)                              & 19                   & 223586               & 120                  \\ \hline
DenseED  & (3,6,3)                              & 19                   & 223586               & 144                  \\ \hline
DenseED  & (6,12,6)                             & 31                   & 788046               & 264                  \\ \hline
DenseED  & (8,8,8)                              & 31                   & 727850               & 236                  \\ \hline
DenseED  & (9,18,9)                             & 43                   & 1663176              & 384                  \\ \hline
\href{https://github.com/ND-HowardGroup/Instant-Image-Denoising/tree/master/Python}{U-Net}    & Encoder: 5 layers, Decoder: 5 layers & 18                   & 989712               & 96                   \\ \hline
\href{https://github.com/ND-HowardGroup/Instant-Image-Denoising/tree/master/Python}{DnCNN}   & 17 layers                            & 17                   & 556096               & 64                   \\ \hline
\end{tabular}
\caption{Comparison of different DenseED configurations and number of convolution layers, number of parameters, and maximum feature maps.}
\label{models_compare}
\end{table}
Initial convolutional layer feature maps are set to 48 except for DnCNN, where the feature maps are 64. In the U-Net ML model, the number of feature maps does not double as the network goes deeper. The above-mentioned U-Net configuration cannot provide the super-resolution with our small training dataset. Still, it contains 18 convolutional layers, and the number of parameters is close to a million. Also, In DenseED-(9,18,9) configuration has more dense layers, which provides access to more in-depth features (384) compared to the U-Net ML model and estimates the super-resolution images.

\begin{sidewaystable}[!t]
\setlength{\extrarowheight}{3pt}
\setlength{\tabcolsep}{3pt}
\begin{tabularx}{\textwidth}{|p{20mm}|p{10mm} |p{25mm} |X |X | p{20mm}| X |}
\hline
Paper & Type & Sample & Network & Dataset & Image size & Input and Target \\
\hline
Deep-STORM \cite{DeepSTORM} & FC & \begin{itemize} \item Simulated microtubules \item quantum dots \end{itemize} &  \begin{itemize} \item Encoder: 3 layers \item Decoder: 3 layers \end{itemize} & \begin{itemize} \item Training: \textcolor{red}{\textbf{7000}} \item validation: 3000 \item Test: 3000 \end{itemize} & 208$\times$208  & \begin{itemize} \item ThunderSTORM plugin (localization) \end{itemize} \\\hline
CNN residuals \cite{ayas2020microscopic_fc2} & FC & Blood samples &  20 layers CNN & \begin{itemize} \item Training: \textcolor{red}{\textbf{16000}} \item Testing: a few FOV's \end{itemize} & 48$\times$48  & \begin{itemize} \item Input: 10x,20x Objective \item Target: 40x Objective \end{itemize} \\\hline
GANs (DL to SR) \cite{wang2019deep_gan1} & GANs & BPAE samples, nanobeads &  \begin{itemize} \item G: U-Net (encoder:4, decoder:4 layers) \item D: FC layers \end{itemize} & \begin{itemize} \item Training: \textcolor{red}{\textbf{2000}} for each fluorophore \item validation: 700  \item Testing: a few FOV's \end{itemize} & 1024 $\times$1024  & \begin{itemize} \item Input: 10x Objective \item Target: 20x Objective \item Cross modality conversion (widefield to confocal) \end{itemize} \\\hline
RFGANM \cite{zhang2019high_gan2} & GANs & Fibroblast in mouse brain &  \begin{itemize} \item G: ResNET (16 layers) \item D: FC layers \end{itemize} & \begin{itemize} \item Training: \textcolor{red}{\textbf{1080}} \end{itemize} & 96$\times$96  & \begin{itemize} \item Target: 384$\times$384 (increase FOV by 4 times) \end{itemize} \\\hline
ANNA-PALM \cite{ouyang2018deep_gan3} & GANs & \begin{itemize} \item Microtubules \item Nanopores \end{itemize} & \begin{itemize} \item G: U-Net (encoder:8, decoder:8 layers) \item D: FC layers \end{itemize} & \begin{itemize} \item Training: \textcolor{red}{\textbf{30000}} dense PALM images \end{itemize} & - & \begin{itemize} \item Input: under-sampled PALM \item Target: dense PALM images \end{itemize} \\\hline
\end{tabularx}
\caption{Summary of existing ML super-resolution methods with fluorescence microscopy data. $G$ and $D$ are Generator and Discriminator in GANs, respectively. DL: diffraction-limited, SR: Super-resolution, FOV: field-of-view, CNN: convolutional neural network.}
\label{ML_compare}
\end{sidewaystable}

\bibliography{report} 

\begin{thebibliography}{10}

\bibitem{huszka2019super}
Huszka, G. and Gijs, M.~A., ``Super-resolution optical imaging: A comparison,''
  {\em Micro and Nano Engineering}~{\bf 2},  7--28 (2019).

\bibitem{GSOS}
Zhang, Y., Benirschke, D., Abdalsalam, O., and Howard, S.~S., ``Generalized
  stepwise optical saturation enables super-resolution fluorescence lifetime
  imaging microscopy,'' {\em Biomedical Optics Express}~{\bf 9}(9),  4077--4093
  (2018).

\bibitem{SRRF_Nature}
Gustafsson, N., Culley, S., Ashdown, G., Owen, D.~M., Pereira, P.~M., and
  Henriques, R., ``Fast live-cell conventional fluorophore nanoscopy with
  {I}mage{J} through super-resolution radial fluctuations,'' {\em Nat.
  Commun.}~{\bf 7},  12471 (2016).

\bibitem{instantFLIM}
Zhang, Y., Guldner, I.~H., Nichols, E.~L., Benirschke, D., Smith, C.~J., Zhang,
  S., and Howard, S.~S., ``High-speed, long-term, 4d in vivo lifetime imaging
  in intact and injured zebrafish and mouse brains by instant flim,'' {\em
  bioRxiv}  (2020).

\bibitem{gustafsson2016fast}
Gustafsson, N., Culley, S., Ashdown, G., Owen, D.~M., Pereira, P.~M., and
  Henriques, R., ``Fast live-cell conventional fluorophore nanoscopy with
  {ImageJ} through super-resolution radial fluctuations,'' {\em Nature
  Communications}~{\bf 7}(1),  1--9 (2016).

\bibitem{mannam2020machine}
Mannam, V., Zhang, Y., Yuan, X., Ravasio, C., and Howard, S.~S., ``Machine
  learning for faster and smarter fluorescence lifetime imaging microscopy,''
  {\em Journal of Physics: Photonics}~{\bf 2}(4),  042005 (2020).

\bibitem{DeepSTORM}
Nehme, E., Weiss, L.~E., Michaeli, T., and Shechtman, Y., ``Deep-{STORM}:
  super-resolution single-molecule microscopy by deep learning,'' {\em
  Optica}~{\bf 5}(4),  458--464 (2018).

\bibitem{mannam2020performance}
Mannam, V. and Kazemi, A., ``Performance analysis of semi-supervised learning
  in the small-data regime using {VAE}s,'' {\em arXiv preprint
  arXiv:2002.12164}  (2020).

\bibitem{osokin2017gans}
Osokin, A., Chessel, A., Carazo~Salas, R.~E., and Vaggi, F., ``Gans for
  biological image synthesis,'' in [{\em Proceedings of the IEEE International
  Conference on Computer Vision}{\nolinebreak\hspace{0.1em}]},   2233--2242
  (2017).

\bibitem{zhu2018bayesian}
Zhu, Y. and Zabaras, N., ``Bayesian deep convolutional encoder--decoder
  networks for surrogate modeling and uncertainty quantification,'' {\em
  Journal of Computational Physics}~{\bf 366},  415--447 (2018).

\bibitem{long2015fully}
Long, J., Shelhamer, E., and Darrell, T., ``Fully convolutional networks for
  semantic segmentation,'' in [{\em Proceedings of the IEEE conference on
  computer vision and pattern recognition}{\nolinebreak\hspace{0.1em}]},
  3431--3440 (2015).

\bibitem{ronneberger2015u}
Ronneberger, O., Fischer, P., and Brox, T., ``U-net: Convolutional networks for
  biomedical image segmentation,'' in [{\em International Conference on Medical
  image computing and computer-assisted
  intervention}{\nolinebreak\hspace{0.1em}]},   234--241, Springer (2015).

\bibitem{mannam2020instant}
Mannam, V., Zhang, Y., Zhu, Y., and Howard, S., ``Instant image denoising
  plugin for {ImageJ} using convolutional neural networks,'' in [{\em
  Microscopy Histopathology and Analytics}{\nolinebreak\hspace{0.1em}]},
  MW2A--3, Optical Society of America (2020).

\bibitem{huang2017densely}
Huang, G., Liu, Z., Van Der~Maaten, L., and Weinberger, K.~Q., ``Densely
  connected convolutional networks,'' in [{\em Proceedings of the IEEE
  conference on computer vision and pattern
  recognition}{\nolinebreak\hspace{0.1em}]},   4700--4708 (2017).

\bibitem{jegou2017one}
J{\'e}gou, S., Drozdzal, M., Vazquez, D., Romero, A., and Bengio, Y., ``The one
  hundred layers tiramisu: Fully convolutional densenets for semantic
  segmentation,'' in [{\em Proceedings of the IEEE conference on computer
  vision and pattern recognition workshops}{\nolinebreak\hspace{0.1em}]},
  11--19 (2017).

\bibitem{ayas2020microscopic_fc2}
Ayas, S. and Ekinci, M., ``Microscopic image super resolution using deep
  convolutional neural networks,'' {\em Multimedia Tools and Applications}~{\bf
  79}(21),  15397--15415 (2020).

\bibitem{wang2019deep_gan1}
Wang, H., Rivenson, Y., Jin, Y., Wei, Z., Gao, R., G{\"u}nayd{\i}n, H.,
  Bentolila, L.~A., Kural, C., and Ozcan, A., ``Deep learning enables
  cross-modality super-resolution in fluorescence microscopy,'' {\em Nature
  Methods}~{\bf 16}(1),  103--110 (2019).

\bibitem{zhang2019high_gan2}
Zhang, H., Fang, C., Xie, X., Yang, Y., Mei, W., Jin, D., and Fei, P.,
  ``High-throughput, high-resolution deep learning microscopy based on
  registration-free generative adversarial network,'' {\em Biomedical Optics
  Express}~{\bf 10}(3),  1044--1063 (2019).

\bibitem{ouyang2018deep_gan3}
Ouyang, W., Aristov, A., Lelek, M., Hao, X., and Zimmer, C., ``Deep learning
  massively accelerates super-resolution localization microscopy,'' {\em Nature
  Biotechnology}~{\bf 36}(5),  460--468 (2018).

\end{thebibliography}
\bibliographystyle{spiebib} 

\end{document}